\documentclass[aps,prl,twocolumn, showpacs]{revtex4}

\usepackage{epsfig}
\usepackage{graphicx}

\begin{document}
\title{Optical Read-Out and Initialization of an Electron Spin in a Single Quantum Dot}

\author{A. Shabaev$^a$, Al. L. Efros$^*$$^a$, D. Gammon$^a$, and I. A. Merkulov$^b$}
\affiliation{$^a$Naval Research
Laboratory, Washington, DC 20375, USA\\
$^b$A. F. Ioffe Physico-Technical Institute, RAS, St. Petersburg, 194021 Russia
}

\begin{abstract}
We describe theoretically the resonant optical excitation of a trion with circularly polarized light and discuss how this trion permits the read-out of a single electron spin through a recycling transition.
Optical pumping through combination of circularly polarized optical  $\pi$--pulses with permanent or $\pi$-- pulsed transverse magnetic fields suggests feasible protocols for spin initialization.
\end{abstract}
\pacs{78.67.Hc,03.67.Lx,78.47.+p,71.35.-y }

\maketitle

An electron spin in a single quantum dot (QD) could be used as a solid state qubit if certain requirements are met\cite{LossDiVincenzo,Imamoglu,Sham}. For example, it must be possible to coherently manipulate single and entangled spins in a time much faster than the spin coherence time and to read out the state of the single electron spin during the spin relaxation time. It has been suggested \cite{LossDiVincenzo} that electronic gates could be used to accomplish these tasks in semiconductor materials, and great effort currently is being exerted to demonstrate the needed quantum tools in nano-structured systems. One of the most difficult of these requirements in a fully electronic implementation is the measurement of the state of a single spin.

Here we consider an alternative optical approach to this problem. If a single unpaired electron is contained in a QD in which an exciton can be excited, the selectivity and sensitivity of optical techniques provides the opportunity to access the spin states in individual and pairs of semiconductor QDs. An optical based computation scheme is of special interest because of the speed that operations could be carried out using ultrafast laser technology, and because of the power of optical selection rules to measure and control spin. Similar techniques have achieved remarkable success in the measurement and control of the hyperfine states of individual ions in magnetic traps for quantum computing \cite{AQC}.  Over the last decade, materials development based on atomic layer growth, doping, and gating techniques have led to semiconductor QD systems in which a single electron can be injected into a QD, and most importantly, there have recently been a number of demonstrations in which a single QD with a single electron spin has been probed optically \cite{Tischler}. Furthermore, several groups have demonstrated optical Rabi oscillations in a single QD \cite{Steel}.
In this letter we demonstrate theoretically that the resonant optical excitation of the spin to an intermediate trion state can lead both to readout and initialization capability in spite of random hyperfine fields of nuclei acting on the electron spin.

We consider singly charged QDs in which the lateral dimensions are much larger than the height - which serves as the quantization axis, $z$, for the electron and hole spin projections \cite{shape}. The ground electron and hole states in these QDs are degenerate with respect to their spin projection $\pm 1/2$ (denoted as $|\uparrow\rangle$ and $|\downarrow\rangle$) and $\pm 3/2$ (denoted as  $|\Uparrow\rangle$ and $|\Downarrow\rangle$ ) respectively. The optical excitation of the charged dots leads to the formation of a trion that consists of two electrons sitting at the same quantum level and a hole (see Fig.1). The $|\uparrow\downarrow\Uparrow\rangle$  ($|\uparrow\downarrow\Downarrow\rangle$) ground state of a trion created  by $\sigma^+$ ($\sigma^-$) polarized light  from the $|\uparrow\rangle$ ($|\downarrow\rangle$)) electron state \cite{Tischler} can decay spontaneously  only into the same initial electron state (see Fig. 1).  This decay process is accompanied by emission of circularly polarized light in which $\sigma^+$ or $\sigma^-$ polarization is uniquely determined by the $|\uparrow\rangle$ or  $|\downarrow\rangle$ electron spin projection, respectively.
Importantly, the selection rules allow the repetition of this cycle many times without losing the electron spin state.

\begin{figure}[th]
\vskip-0.1truecm
\begin{center}
\epsfig{file=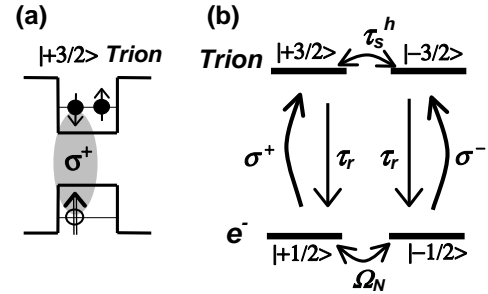, width=0.35\textwidth}
\end{center}
\vskip-0.7truecm
\caption{ Schematic presentation of (a): the trion created by $\sigma^+$ polarized light in singly charged QDs; (b): coherent \& spontaneous optical transitions and spin relaxation processes that control the spin dynamics in the singly charged QD.}
\label{level}
\end{figure}
\vskip-0.1truecm

As a result the polarization of emitted light allows us, in principle, to measure the electron spin projection through this recycling transition. However the electron and trion spin relaxation will limit the measurement time of the photoluminescence (PL) with the given polarization.  It is obvious that one would want to collect as many photons as possible from a single QD before the electron spin relaxes. Even under resonant excitation condition, however, the rate of emitted photons can not exceed $1/\tau_r$, where $\tau_r$ is the trion radiative decay time. At first glance this limits significantly the possibility of optical measurements of the spin state in a single QD because the spin relaxation time does not exceed $\sim 100\tau_r$ even in very optimistic estimates. 

Electron spin relaxation at low temperatures is mainly caused by its interaction with nuclei. In a zero external magnetic field the electron spin precesses in the hyperfine field, $\mbox{\boldmath $B$}_N$, of the frozen configuration of nuclear spins interacting with the localized electron  with frequency $\Omega_N=\mu_Bg_eB_N/\hbar$ \cite{Merkulov}, where $\mu_B$ is the Bohr magneton, and $g_e$ is the electron $g$-factor.  $\mbox{\boldmath $B$}_N$ has an arbitrary direction and its component perpendicular to the $z$ axis of the QD leads to electron spin relaxation \cite{spinrelax}.  Calculation shows that the electron spin relaxation time  is on the order of $\sim 1-10$\,ns, depending on the QD size \cite{Merkulov}. This  is usually longer than the radiative decay time of  the trion \cite{trion}. In contrast, the trion should not interact strongly with the nuclear spin because its electrons are in a singlet state and a hole has negligable hyperfine interaction. The hole spin relaxation in the trion state is a two--phonon assisted process \cite{Takagahara}. The probability of such processes is proportional to the phonon concentration, which decreases exponentially with temperature.
As a result the trion spin relaxation time, $\tau_s^h$ (as well as the electron spin time, $\tau_s^e$, connected with phonon assisted transitions) can be much longer than the electron spin relaxation time by nuclear interactions at low temperatures, and it is  the latter that should limit our ability to measure the electron spin orientation.  We will show, however,  that electron spin relaxation in a nuclear field is suppressed by  light in the strong coupling regime, and as a result the electron spin state in a single QD can be measured optically. 

Intense resonant light drives the transitions between the electron and trion states. The frequency of these population oscillations is given by the Rabi frequency ($\Omega_R$) at resonant excitation,  $\Omega_R=2dE/\hbar$, where $d$ is the dipole moment of the optical transition and $E$ is the electric field amplitude.  Under strong excitation conditions when $\Omega_R\gg \Omega_N$ the electron spin can not significantly change its direction because the spin precession is interrupted by the light stimulated excitation of the electron into the trion state.  The spin precession is interrupted periodically at the Rabi frequency, and at high excitation intensity the spin can change its projection by only a very small value $(\Omega_N/ \Omega_R)^2\ll 1$, a type of "line narrowing" \cite{OO}.  As a result spin relaxation arising from the hyperfine field is suppressed. 

The spontaneous decay of the trion is the incoherent process that destroys the synchronization of the electron spin motion with electromagnetic excitation and leads finally to electron spin relaxation in the strong excitation regime. A strong, polarized laser field splits the electron $\pm 1/2$ spin states because of Rabi splitting.  This reduces the admixture of the -1/2 spin state to the +1/2 spin state, which is proportional now to the small value $\Omega_N/ \Omega_R$.    As a result the rate of spontaneous trion decay into the electron -1/2  spin state that describes the electron spin relaxation  can be obtained in second order perturbation theory:
\begin{eqnarray}
{1\over T_s^e}={1\over \tau_r}\left({\Omega_N \over \Omega_R}\right)^2
\end{eqnarray}
The time of electron spin relaxation $T_s^e$ under strong excitation, $\Omega_N/ \Omega_R\ll 1$, is much longer than $\tau_r$. Thus, the optical recycling transition provides more PL photons before relaxation occurs.

\begin{figure}[th]
\vskip-1truecm
\begin{center}
\epsfig{file=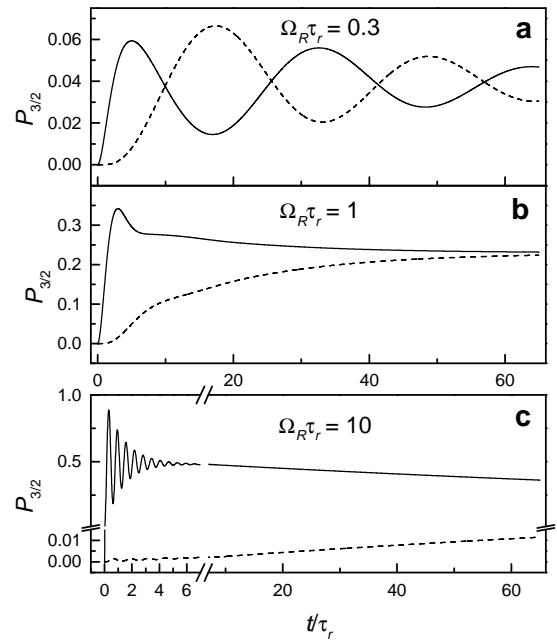, width=0.45\textwidth}
\end{center}
\vskip-2truecm
\caption{Time dependence of the $|\uparrow\downarrow\Uparrow\rangle$ trion state population, $P_{3/2}$, for three excitation intensities for $\Omega_N\tau_r=0.1$. The solid and dashed lines are calculated for the $|\uparrow\rangle$ and $|\downarrow\rangle$ initial electron spin state, respectively.}
\label{timepho}
\end{figure}
\vskip-0.2truecm

We explore this problem completely using the density matrix approach for the four level system of Fig.1b. Polarized  $\sigma^+$ light with frequency $\omega$ connects  only the electron $+1/2$ spin state with the trion $+3/2$ spin state.  Electron spin relaxation arising from the hyperfine field  leads to transitions between the electron spin states. We include now the hole spin relaxation, which leads to transitions between the trion spin states. 

Figure 2 shows the time dependence of the diagonal component of the density matrix  $P_{3/2}$ describing the population of the $+3/2$ trion state. $P_{3/2}$ characterizes the intensity of the $\sigma^+$ polarized PL because the rate of the $\sigma^+$ polarized photon emission $\sim P_{3/2}/\tau_r$.    The calculations were done for the two initial electron spin orientation ($s_z=\pm 1/2$) for zero detuning, $\varepsilon_{3/2} - \varepsilon_{1/2} -\hbar\omega = 0$.  One can see that at low excitation intensity (see Fig. 2a) the population of the trion state $P_{3/2}$ excited by $\sigma^+$ polarized light oscillates with the frequency of the electron spin precession in the hyperfine field of the nuclei. The oscillations have opposite phase for the two opposite initial conditions of the electron spin.  At intermediate excitation intensity $\Omega_R\tau_r=1$ (see Fig. 2b) the precession oscillations are suppressed and the $\sigma^+$ polarized PL from the QD when the electron initially has +1/2 spin is more intense  then if the electron has -1/2 spin. When the optical excitation is further increased (Fig. 2c), Rabi oscillations appear at short times and the difference in PL intensity for the two cases remains much longer. The difference in the PL intensity for the two cases allow us to measure the electron spin polarization with high accuracy in the case of strong optical excitation (see Fig. 2c).
The technically challenging problem to distinguish emission from the scattering light in this scheme can be resolved for example by using time resolved technique.

It is important to note that eventually the density matrix relaxes to the stationary state, which does not depend on the initial condition. That is why only the transient part of the PL polarization, measured at $t < T_s^e$, provides information on the electron spin polarization. At longer times {\em optical pumping} of the electron spin becomes important.
This suggests using intense circularly polarized light for optical {\em initialization} of the electron spins in the QDs.

Optical pumping of electron spins was demonstrated in atoms almost 50 years ago in a seminal paper of Brossel and Kastler \cite{Kastler}. In those experiments circularly polarized light  transfered an electron from one spin sublevel into an excited state, which then decayed spontaneously  into both spin sublevels of the ground state. 
This leads to the depletion of one of the ground state sublevels and accumulation in the other spin sublevel. Unfortunately, this optical pumping scheme does not normally work  in our QDs \cite{shape} because the excited trion decays only into the same initial state (see Fig. 1a). 

\begin{figure}[th]
\vskip-0.8truecm
\begin{center}
\epsfig{file=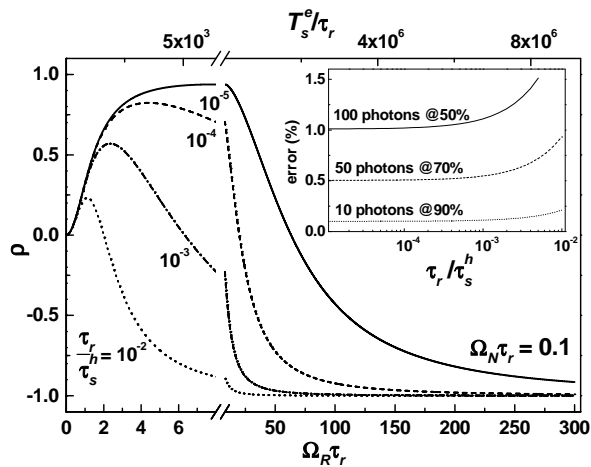, angle=-90, width=0.5\textwidth}
\end{center}
\vskip-0.7truecm
\caption{The dependence of the steady state degree of electron spin polarization on $\Omega_R$ that is proportional to the laser field amplitude for different hole spin relaxation rates, $1/\tau_s^h$. Inset: The dependence of the single spin measurement error on the hole spin relaxation rate, $\Omega_R\tau_r = 10$.}
\label{sizegfact}
\end{figure}
\vskip-0.2truecm

Electron and hole spin relaxation allows us to overcome this restriction. Figure 3 shows the degree of electron spin polarization $\rho=(P_{1/2}-P_{-1/2})/(P_{1/2}+P_{-1/2})$   calculated as a function of excitation intensity in steady state ($t\rightarrow \infty$), where $P_{\pm 1/2}$ are the populations of the electron states.  At times $t > T_s^e$ the intense $\sigma^+$ polarized light  keeps the electron in one of its  $\pm 1/2$ spin states independent of the initial electron spin orientation. In the case when $T_s^e\ll \tau_s^h$  one can completely neglect the hole spin relaxation processes.  The steady state solutions of the density matrix equations show that under strong excitation conditions, $\Omega_R\gg 1/\tau_r \gg \Omega_N$, an electron is practically always in the +1/2 spin state and  the probability to find it in the -1/2 spin state is proportional to $1/(\tau_r\Omega_R)^2 \ll 1$. In the opposite limiting case, $T_s^e\gg \tau_s^h$, as a result of the hole spin flip processes, the electron is finally transfered to the optically passive -1/2 state similar to that in the pumping experiment of Ref. \onlinecite{Kastler}. The probability to find an electron in the +1/2 spin state is  $\sim \tau_s^h/T_s^e\ll 1$. 

As one can see in Fig. 3, a high degree of electron spin polarization  can be achieved using a very intense polarized light source, such that $\Omega_R\tau_r>>1$. However, optical pumping of electron spin can also be achieved using a remarkable property of our QDs: the trion state is not affected by a  magnetic field lying in the quantum wall (QW) plane in a first approximation. This allows one to use a combination of $\pi$--optical pulses and a transverse magnetic field for spin initialization in QDs.

The transverse magnetic field does not affect the singlet trion, consisting of  two electrons in a singlet state and a heavy hole, because the  $|\Uparrow\rangle$ and $|\Downarrow\rangle$  heavy hole states have zero g-factor in the plane of the quantum well \cite{IvchenkoPikus}.  A simple protocol leading to electron spin orientation is a sequential action of optical and magnetic $\pi$--pulses. The first $\sigma^+$ ($\sigma^-$) polarized optical $\pi$--pulse creates the $|\uparrow\downarrow\Uparrow\rangle$  ($|\uparrow\downarrow\Downarrow\rangle$) singlet trion state if the QD contains an electron in the $|\uparrow\rangle$ ($|\downarrow\rangle$) spin state.  According to the selection rules, this optical pulse does not affect
the QD if the electron is in the opposite  $|\downarrow\rangle$ ($|\uparrow\rangle$) spin state.  The electron in this QD can be transfered to the $|\uparrow\rangle$ ($|\downarrow\rangle$) spin state by the short magnetic $\pi$--pulse which at the same time would not affect the trion state, if it was formed.  As a result 100\% polarized electrons can be prepared in the $|\uparrow\rangle$ ($|\downarrow\rangle$) spin state after the trion relaxation to the ground state of the QD.  For realization of this protocol the duration of the magnetic $\pi$--pulse should be shorter than the lifetime of the trion\cite{trion}. Creation of short magnetic pulses is a difficult technical problem that perhaps can be solved, for example, using electric field variation of the electron g-factor in a permanent magnetic field \cite{Yablonovitch,Sales}.
 
 Alternatively, the combined effect of $\pi$--optical pulses and {\em permanent} transverse magnetic field  allows one to initialize an electron spin into a state with a well defined phase.
The permanent magnetic field, $B$, lying in the QW plane  leads to the precession of the electron spin with Larmor frequency $\Omega_e=\mu_Bg_eB/\hbar$.  However, the phase of its precession is not defined. The phase of the electron spin precession can be fixed by applying synchronized periodic optical $\pi$--pulses with period $T_p = 2\pi n/\Omega_e$, where $n=1,2,...$. Such an experimental setup was used already by Kikkawa and Awschalom  for nonresonant excitation of QWs \cite{Kikkawa}. 

To describe this process quantitatively, let us consider the dynamics of the electron and trion spin polarization during the time, $t$, between two optical circularly polarized $\pi$--pulses: $t_{N+1}>t>t_N$, where $t_N=NT_p$ ($N=0,1,...$). In vector representation, the evolution of the electron spin polarization $\mbox{\boldmath $S$}$ is described by:
\begin{equation}
{d \mbox{\boldmath $S$}\over dt}=[\mbox{\boldmath $\Omega$}_e\times \mbox{\boldmath $S$}]-{\mbox{\boldmath $S$}\over \tau_s^e}+{J(t)\over \tau_r}\mbox{\boldmath $e$}_z~,
\end{equation}
where $\mbox{\boldmath $e$}_z$ is the unit vector along the $z$ axis, $\mbox{\boldmath $\Omega$}_e\perp\mbox{\boldmath $e$}_z$, $\tau_s^e$ is the electron spin relaxation time, which is much longer than $2\pi/\Omega_N$ in an external magnetic field $B\gg B_N$ \cite{OO}.  $J(t)=J(t_N)\exp(-\gamma \Delta t)$ is the time dependence of 1/3 of the average  spin projection of the trion hole on the $z$-- axis after the optical $\pi$--pulse, where $\Delta t=t-t_N$ and $\gamma=1/\tau_r+1/\tau_s^h$. Equation (2) can be solved for the  complex spin components $S_{\pm} =S_y\pm iS_z$ representing rotation of the spin around the magnetic field, which is taken along the $x$--axis. 

The important characteristic of the electron spin precession that describes electron spin polarization along the $z$-axis is  $S_z(t)= {\rm Im}[S_+(t)]$.  At the time when the trion recombines to the ground one--electron state ($\Delta t \gg \tau_r$), $S_z(t)=0.5[P_{1/2}(t)-P_{-1/2}(t)]$,  
 and its time dependence can be written
\begin{equation}
S_z(t)={\rm Re}\left[\left(S_z(t_N+\delta)+{J(t_N+\delta)\over \tau_r(\gamma+i\omega)}\right)e^{i\omega \Delta t}\right],
\end{equation}
 where $\omega=\Omega_e+ i/\tau_s^e$ and $\delta\rightarrow 0$.  The initial values of the electron and trion spin polarizations in Eq.3  after the $\sigma^+$ polarized $\pi$--pulse  are connected with the density matrix elements before this pulse: $S_z(t_N+\delta)=-0.5P_{-1/2}(t_N-\delta)$ and $J(t_N+\delta)=0.5P_{1/2}(t_N-\delta)$. These values are expressed through the electron spin polarization  $P_{1/2}(t_N-\delta)=0.5[2S_z(t_N-\delta)+1]$ and $P_{-1/2}(t_N-\delta)=0.5[1-2S_z(t_N-\delta)]$. Substituting $S_z(t_N+\delta)$ and $J(t_N+\delta)$ into  Eq.(3)  one can obtain the recursive relation for $S_z(t_{N+1}-\delta)$ and $S_z(t_N-\delta)$:
\begin{eqnarray}
S_z(t_{N+1})={\rm Re}\left\{\left[{2S_z(t_N)-1\over 4}+{2S_z(t_N)+1\over 4\tau_r(\gamma+i\omega)}\right]e^{i\omega T_p}\right\}.\nonumber
\end{eqnarray}

Let us consider the spin dynamics of the initially unpolarized electron spin  after action of the $\sigma^+$--polarized $\pi$--pulse. In this case the initial condition: $S_z(0-\delta)=0$. The recursive relation has a clear physical interpretation in two limiting cases. First, when $\Omega_e\tau_r\ll 1$   the denominator $\tau_r(\gamma+i\omega)\approx 1$ because $\tau_s^h,\tau_s^e\gg \tau_r$. The electron spin polarization is conserved during trion relaxation; the trion contribution to the electron spin polarization, $1/4$, completely compensates the electron contribution, $-1/4$, and the electron remains unpolarized.  In the opposite limit of strong magnetic fields,  $\Omega_e\tau_r\gg 1$, the electron left after slow recombination of the trion is out of phase and this compensation does not happen. As a result, a significant electron spin polarization is created in the QD after trion recombination.

\begin{figure}[th]
\vskip-1.5truecm
\begin{center}
\epsfig{file=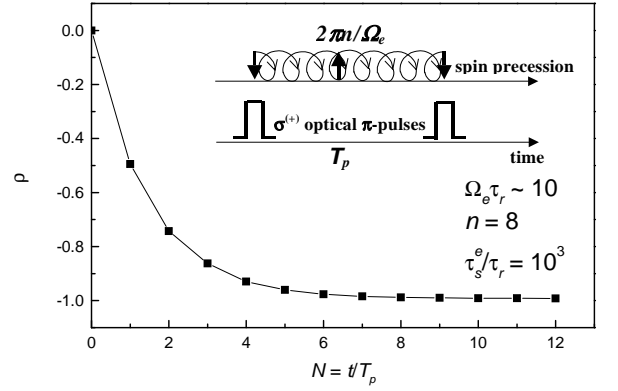, angle=-90, width=0.5\textwidth}
\end{center}
\vskip-0.8truecm
\caption{The degree of electron spin polarization  as a function of the $\sigma^+$ polarized optical $\pi$--pulse number ($N$).}
\label{pumping}
\end{figure}
\vskip-0.3truecm

Repeating the optical $\pi$--pulses allows us to pump the electron spin of a certain phase to high  polarization if
 the electron spin relaxation time is long enough: $T_p\ll \tau_s^e$.  Fig. 4 shows the degree of  electron spin polarization, $\rho(t_N)=2S_z(t_N)$, as a function of the $\pi$--pulse number. The polarization degree reaches 99\% after $N=8$ pulses. 

This last {\em initialization} protocol  can  be matched with the {\em read out} scheme suggested above. Intense resonant steady state excitation of the trion by circularly polarized light turned on at $t=NT_p$ would allow one to measure the electron spin polarization if $\Omega_R\gg \Omega_e$.

The above consideration assumed ideal selection rules which hold  for QDs with cylindrical shape or at least $C_{2v}$ symmetry. Only significant deviations from this 
shape can change them in epitaxial QDs due to weak admixture of light and heavy hole subbands. To check the feasibility of our measurement scheme we include weakly allowed forbidden transitions in our four level model as well as additional dephasing which may be connected with fluctuations of the trion-electron transition energy in a random electric field. These deviations from the ideal model increase the spin relaxation rate. They do not affect significantly, however,  our ability to determine an electron spin by measuring the burst of the photons or its absence.  The effect can be quantified by the measurement error introduced in \cite{Calarco}. The inset in Fig. 3 shows the dependence of the error on the total hole spin relaxation rate where both additional mechanisms are included into $1/\tau_s^h$ and $1/T_s^e$, calculated for various photon bursts and registration efficiencies 50\%, 70\% and 90\%. An additional dephasing rate  smaller than $1/\tau_r$ and  probability of the forbidden transition smaller then $0.001/\tau_r$ does not practically increase the error of our measurements.

In conclusion, our analysis shows that polarization of a single electron spin in a QD can be measured optically using an intense resonant light source that suppresses electron spin relaxation.  We also propose using a combination of $\pi$-pulses and transverse magnetic field for optical pumping of the electron spins in  QDs. The simplest protocol  leading to electron spin orientation  is a sequential combination of optical and magnetic $\pi$-pulses. A high degree of electron spin polarization can  also be achieved with a permanent transverse magnetic field  by applying optical $\pi$- pulses with repetition rate equal to an integer multiple of the magnetic precession frequency.

{\em Acknowledgments,} the authors acknowledge financial support from DARPA, ONR and NSA/ARDA. In addition  A. S. thanks the NRC  and I. A. M. thanks  CRDF for support.


\begin{thebibliography}{99}
\bibitem[*]{address} All correspondence should be directed to efros@dave.nrl.navy.mil.
\bibitem{LossDiVincenzo} D. Loss, and D. P. DiVincenzo,  Phys. Rev. A {\bf 57}, 120 (1998).
\bibitem{Imamoglu} A. Imamo\={g}lu {\em et al.}, Phys. Rev. Lett. {\bf 83}, 4204 (1999).
\bibitem{Sham} C. Piermarocchi {\em et al.}, Phys. Rev. Lett. {\bf 89}, 167402 (2002).
\bibitem{AQC} C. Monroe, {\em Nature} {\bf 416}, 238 (2002).
\bibitem{Tischler} A. Hartmann {\em et al.}, Phys. Rev. Lett. {\bf 84}, 5648 (2000);
R.J. Warburten {\em et al.}, Nature {\bf 405}, 926 (2001);
J. G.  Tischler {\em et al.},  Phys. Rev. B {\bf 66}, 081310  (2002).
\bibitem{Steel} T. H. Stievater {\em et al.}, Phys. Rev. Lett. {\bf 87}, 133603 (2001);
 H. Kamada et al., Phys. Rev. Lett. {\bf 87}, 246401 (2001); H. Htoon et al., Phys. Rev Lett. {\bf 88}, 087401 (2002);
 A. Zrenner et al., Nature {\bf 418}, 612 (2002).
\bibitem{shape} For example, QDs self assembled in molecular beam epitaxy have this symmetry.
\bibitem{Merkulov}  I. A. Merkulov, Al. L. Efros, and M. Rosen, Phys. Rev. B {\bf 65}, 205309 (2002).
\bibitem{spinrelax}Electron spin precession in a hyperfine nuclear field, although generally not an irreversible process, leads to electron spin relaxation here in the presence of spontaneous emission.  
\bibitem{trion} In natural GaAs QDs $\tau_r\sim 100$\,ps, see for example: G. Finkelstein {\em et al.}, Phys. Rev. B {\bf 58}, 12637 (1998).
\bibitem{Takagahara}  T. Takagahara, Phys. Rev. B {\bf 62}, 16840 (2000); we estimate $\tau_s^h \sim 1 \mu s$ at $T = 4 K$.
\bibitem{OO}{\em Optical Orientation}, edited by B. Meier and B. P. Zakharchenya (North Holland, Amsterdam, 1984), Ch.II.
\bibitem{Kastler}J. Brossel and A. Kastler, Comp. Rend. {\bf 229}, 1213 (1949).
\bibitem{IvchenkoPikus} V. F. Sapega {\em et al.}, Phys. Rev. B 45, 4320 (1992).
\bibitem{Yablonovitch}H. W. Jiang and Eli Yablonovitch, Phys. Rev. B {\bf 64}, 041307(R) (2001).
\bibitem{Sales} G. Sales {\em et al.}, Nature {\bf 414}, 619 (2001)
\bibitem{Kikkawa} J. M. Kikkawa and D. D. Awschalom, Science {\bf 287}, 473 (2000).
\bibitem{Calarco} T. Calarco {\em et al.}, quant-ph 00304044.
\end{thebibliography}
\end{document}